# Conduction electrons in acceptor-doped GaAs/GaAlAs heterostructures: a review

Wlodek Zawadzki[*,1], Andre Raymond[2], and Maciej Kubisa[3]

[1]Institute of Physics, Polish Academy of Sciences, 02-668 Warsaw, Poland

[2]Laboratoire Charles Coulomb (LC2) UMR 5221, CNRS, Universite de Montpellier, 34095 Montpellier, France

[3]Laboratory for Optical Spectroscopy of Nanostructures, Department of Experimental Physics, Wroclaw University of Technology, 50-370 Wroclaw, Poland

[*]Corresponding author: zawad@ifpan.edu.pl

ABSTRACT: We review magneto-optical and magneto-transport effects in GaAs/GaAlAs heterostructures doped in GaAlAs barriers with donors, providing two-dimensional electron gas (2DEG) in GaAs quantum wells, and additionally doped with smaller amounts of acceptors (mostly Be atoms) in the vicinity of 2DEG. One may also deal with residual acceptors (mostly C atoms). Behavior of such systems in the presence of a magnetic field differs appreciably from those doped in the vicinity of 2DEG with donors. Three subjects related to the acceptor-doped heterostructures are considered. First is the problem of bound states of conduction electrons confined to the vicinity of negatively charged acceptors by a joint effect of a quantum well and an external magnetic field parallel to the growth direction. A variational theory of such states is presented, demonstrating that an electron turning around a repulsive center has discrete energies above the corresponding Landau levels. An experimental evidence for the discrete electron energies comes from the work on interband photo-magneto-luminescence, intraband cyclotron resonance and quantum magneto-transport (the Quantum Hall and Shubnikov–de Haas effects). An electron rain down effect at weak electric fields and a boil off effect at strong electric fields are introduced. It is demonstrated both theoretically and experimentally that a negatively charged acceptor can localize more than one electron. The second subject describes experiment and theory of asymmetric quantized Hall and Shubnikov–de Haas plateaus in acceptor-doped GaAs/GaAlAs heterostructures. It is shown that main features of the plateau asymmetry can be attributed to asymmetric density of Landau states in the presence of acceptors. However, at high magnetic fields also the rain down effect is at work. The third subject treats the so called disorder modes (DMs) in the cyclotron resonance of conduction electrons. The DMs originate from random distributions of negatively charged acceptor ions whose potentials provide effective quantum wells trapping the conduction electrons. This results in an upward energy shift of DM as compared to the cyclotron resonance. Theory and experimental characteristics of DMs are discussed. A similarity between acceptor-doped heterostructures and 2D systems with antidots is briefly described. In conclusions, we mention weaker points in the research on acceptor-doped heterostructures and indicate possible subjects for further investigations. An effort has been made to quote all important works on the subject.

1. INTRODUCTION

We review magneto-optical and magneto-transport properties of conduction electrons in acceptor-doped GaAs/GaAlAs heterostructures. The subject may appear surprising since acceptors have usually little influence on the conduction electrons in semiconductors. Thus, we should properly define our system in order to avoid possible misunderstandings. We will not be interested in p-type heterostructures. Instead, we will consider heterostructures doped in the GaAlAs barriers with donors (mostly Si atoms), which provide two-dimensional (2D) conduction electrons in GaAs quantum wells (QWs). In addition, such structures can be doped in smaller quantities in the GaAs well by either donors or acceptors, in order to influence the 2D electrons, and it is the acceptor doping that is of our interest. It turns out that donor-doped or acceptor-doped heterostructures behave quite differently in the presence of a magnetic field. Since the potential wells in the considered systems contain 2D electrons coming from the donors in the barrier, the additional acceptors in smaller quantities will be filled with electrons, i.e. they will represent negatively charged centers, repulsive for the conduction electrons.

The first question that may arise is: how will such repulsive centers influence the conduction electrons in the Landau levels (LLs). The standard answer is: the repulsive centers will broaden LLs on the higher energy sides, the same way the attractive centers broaden LLs on the lower energy sides. This answer is correct but not complete. It was predicted that, if a conduction electron moves in the field of a repulsive center in the presence of a magnetic field parallel to the growth direction, it will have *discrete repulsive energies* above LLs. The first part of our review presents the theory of this phenomenon and its experimental confirmations in magneto-optical and magneto-transport investigations. We will call the charged acceptors in the presence of a magnetic field "magneto-acceptors" (MA) being aware that this term may be somewhat misleading, as it may suggest that we are interested in magnetic properties of acceptors. This is not the case. The system of a charged acceptor center localizing a conduction electron will be marked $A^{1-}$. We will also show that a negatively charged acceptor ion can localize more than one conduction electron.

The second part of the review is concerned with asymmetric broadening of the conduction Landau levels in heterostructures doped with acceptors in the above sense. This property was first observed experimentally in the Quantum Hall Effect and the Shubnikov-de Haas Effect and then described theoretically with the use of asymmetric density of states. In this part we will quote, as a matter of exception, also some results for donor-doped heterostructures in order to emphasize the differences between the two situations. Finally, the third part of the review is concerned with effects of disorder introduced by random distributions of acceptors. The influence of disorder was first observed in the cyclotron resonance (CR) of electrons in Si-MOS structures but found its development in acceptor-doped GaAs/GaAlAs heterostructures. At the end we briefly treat a similarity between the acceptor-doped heterostructures and antidots embedded in two-dimensional electron gas. We will mention main historical points when presenting each of the three parts. In fact, the presentation of each part follows roughly the chronological order of development.

As to the balance between experiment and theory, our review is somewhat biased toward experimental facts, while the theories are presented mostly by their results rather than derivations. Readers interested in theoretical details can consult the original papers. One should add that there exist not too many theoretical papers on the subjects of our interest. Our main purpose is to unite the three different aspects of the acceptor problem and to put into order various notions in each of the subjects. An effort has been made to quote the relevant literature on acceptor-doped heterostructures. We have also tried to present efforts of the important research groups by quoting their main results in the form of figures.

As follows from our review, there still exist gaps and non-understood points in the physics of acceptor-doped heterostructures. The overall feeling is that the coverage of some aspects is "thin" and requires additional confirmation or disproval. There are still unexplored possibilities of acceptor-doped heterostructures with their special features. We hope that the review will motivate further efforts to bridge the gaps, clarify the weak points and apply in practice acceptor-doped two-dimensional systems.

## 2. DISCRETE STATES OF CONDUCTION ELECTRONS LOCALIZED BY NEGATIVE ACCEPTOR IONS

It is known that, in semiconductors, donor atoms bind conduction electrons producing electron states below the conduction band, while acceptor atoms bind valence electrons producing electron states above the valence band. This is related to the sign of potential energy, negative for donor ions and positive for acceptor ions. Thus, in the bulk material, a conduction electron is not bound to an acceptor ion since it is repulsed by the Coulomb interaction. However, if a conduction electron is confined to the proximity of a negative acceptor ion, it will form states with energies near the conduction band. A situation of this kind is provided by a quantum well with an external magnetic field transverse to the 2D plane. In such a system the electron may not run away from the acceptor because it is confined in all directions: along the $z$ axis by the potential well and in the $x$-$y$ plane by the Lorentz force related to the magnetic field. Kubisa and Zawadzki [1, 2, 3] showed that a 2D conduction electron, moving in the presence of a magnetic field and potential of a negatively charged acceptor would have *discrete repulsive energies above the conduction Landau states*. One should mention that Laughlin [4] used first order perturbation theory to show that a positive potential in a 2D system gives a discrete level above the conduction Landau state. A related situation was considered by Yang and MacDonald [5].

Before going to details of the problem we show schematically in Fig. 1 that, in two dimensions, a combination of the repulsive Coulomb potential and the parabolic potential related to an external magnetic field gives a potential minimum having circular symmetry. It is clear that conduction electrons in the resulting circular well will have discrete positive energies and localized wave functions. It follows from the scheme that, as the magnetic field is increased, the repulsive energies become higher. Both above conclusions are confirmed by calculations.

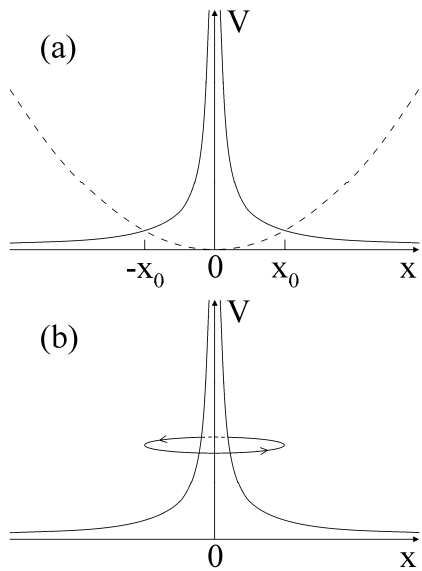

Figure 1. a) Repulsive Coulomb potential of a negatively charged acceptor ion (solid lines) and the parabolic potential of an external magnetic field (schematically). b) Circular cyclotron-like orbit of a conduction electron in the ground state around the repulsive acceptor ion.

Before presenting the theory, we want to say a few words about putting the center of magnetic potential at the acceptor's position, as indicated in Fig. 1. The magnetic field is always introduced that way and this convenient choice is justified by the invariance of the Schrodinger and Dirac equations with respect to the gauge transformations. However, one can justify this choice without any calculations. Suppose we take the gauge of the vector potential in the form $A = \left(-\tfrac{1}{2}By, +\tfrac{1}{2}Bx, 0\right)$. This gives the magnetic field curl $A = B$ along the $z$ direction, as it should. The above choice of $A$ means that we have taken the center of coordinate system at $x = y = 0$. But the field $B$ is the same everywhere, so that observables related to the acceptor may not depend on acceptor's position. Thus we can put the acceptor at $x = y = 0$, and that is what we do. If we take the gauge using $y' = y - a$ and $x' = x - b$, which again gives the field $B$, we can put the acceptor at $r' = (b, a)$ with the same result.

## 2.1 THEORY

In the following we summarize theoretical results of Refs. [1, 2, 3] describing the discrete energies of $A^{1-}$ centers above the Landau levels and magneto-optical transitions in the vicinity of cyclotron resonance. We consider a conduction electron in a magnetic field in a quasi 2D system characterized by the potential $U(z)$. In addition, the electron moves in a potential $V(r)$ provided by either an ionized negative acceptor or a donor. In the one-band effective mass approximation the Hamiltonian reads

$$H = \frac{1}{2m^*}\boldsymbol{P}^2 + U(z) + V(|\boldsymbol{r}|), \qquad (1)$$

where $\boldsymbol{P} = \boldsymbol{p} + e\boldsymbol{A}/c$ is the kinetic momentum and $m^*$ is the effective electron mass. The vector potential is taken in the symmetric form $\boldsymbol{A} = \left(-\tfrac{1}{2}By, +\tfrac{1}{2}Bx, 0\right)$. To solve the eigenvalue problem one uses the variational method, first reducing the 3D equation to an effective 2D equation. The procedure is based on two approximations. First, the trial function is separated to a product

$$\Psi(\boldsymbol{r}) = \Phi(x, y) f_0(z). \qquad (2)$$

Second, the function $f_0$ is assumed to be the same as that of the electron in the absence of impurity and having the subband energy $\varepsilon_0$. Calculating the variational energy $E$ one obtains the effective 2D equation

$$\left\langle \Phi \left| \frac{1}{2m^*}\left(P_x^2 + P_y^2\right) + V_{eff}(x, y) \right| \Phi \right\rangle = E - \varepsilon_0, \qquad (3)$$

where

$$V_{eff}(x, y) = \int_{-\infty}^{+\infty} f_0^2(z) V(r) \, dz \qquad (4)$$

is the effective 2D impurity potential. Equations (3) and (4) present the desired 2D formulation. One introduces polar coordinates because the system has a cylindrical symmetry. The screening of the Coulomb interaction by 2D electron gas is introduced in the form given by Price [6] and the review of Ando et al. [7]. This approach is valid for the degenerate 2DEG and it does not take into account the effect of magnetic field on the screening.

Eigenfunctions for a free 2D electron in a magnetic field are characterized by two quantum numbers: $N = 0, 1, 2, \ldots$ and the projection of the angular momentum on the magnetic field direction $M = \ldots -2, -1, 0, 1, 2, \ldots$ (spin is omitted). The energies of Landau levels are

$$E_{NM} = 2\gamma\left(N + \frac{M + |M|}{2} + \frac{1}{2}\right), \qquad (5)$$

where $\gamma = \hbar\omega_c/2Ry^*$, $\omega_c = eB/m^*c$ is the cyclotron frequency, $Ry^* = m^*e^4/2\varepsilon^2\hbar^2$ is the effective Rydberg, and $\varepsilon$ is the dielectric constant. The Landau quantum number is $n = N + (M + |M|)/2$. The number $M$ is also a good quantum number in the presence of impurity potential. In order to calculate the energy by a variational procedure one assumes the trial functions to be in the form (see Ref. [3])

$$\Phi_{0M}(\rho,\varphi) = C_{0M}\exp(iM\varphi)\rho^m\exp\left(-\tfrac{1}{4}\alpha_0^2\rho^2 - \beta_0\rho\right), \quad (6)$$

where $C$ is a normalization coefficient, $\alpha_0$ and $\beta_0$ are variational parameters and $m = |M|$. For $N = 0$ and $M \leq 0$ the states $A^{1-}$ are related to the lowest Landau level $n = 0$. The above functions give good results at low as well as high magnetic fields, see Ref. [8]. The employed dielectric constant is $\varepsilon = 12.9$, assuming the same values for GaAs and GaAlAs. This gives $Ry^* = 5.4$ meV.

Figure 2 shows the calculated energies of MA and magneto-donor (MD) states related to the $n = 0$ conduction LL in GaAs/GaAlAs heterostructure as functions of impurity position, and Figure 3 the corresponding energies as functions of the magnetic field for impurities located at the interface. The MA and MD energies related to the $n = 1$ LL are similar to those shown in Fig. 3 with somewhat smaller absolute values, see Ref. [3]. This is a consequence of the fact that the impurity states associated with $n = 1$ are larger than those related to $n = 0$. The impurity energies are counted from the free electron energy of the Landau level. It can be seen that the energies of MA and MD states depend strongly on the impurity position. This dependence is related to the strength of the effective potential in Eq. (4), which is in turn determined by the shape of the envelope function describing the electric subband. As expected, the MA repulsive energies occur above the Landau level, while the MD binding energies occur below it. At $B = 0$ the MA energies coalesce to the free electron energy $E = 0$. This indicates, in agreement with the intuitive expectations, that it is the magnetic field which keeps the conduction electron in the proximity of acceptor ion. In the absence of magnetic field the electron runs away from the acceptor ion and the Coulomb energy vanishes. We emphasize that while the negative (donor) energies are binding energies, the positive (acceptor) ones are repulsive energies.

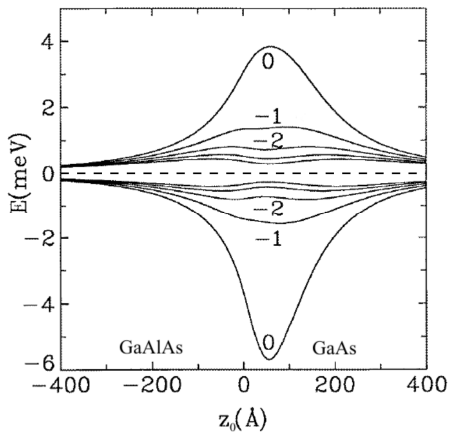

Figure 2. Binding energies of magneto-donor (negative) and repulsive energies of magneto-acceptor states $A^{1-}$ (positive) related to the conduction Landau level $n = 0$ in GaAs/GaAlAs heterojunction, calculated for $B = 10$ T as functions of impurity position. The quantum numbers in both cases are $N = 0$ and $M = 0, -1, -2, ...$ After Ref. [3].

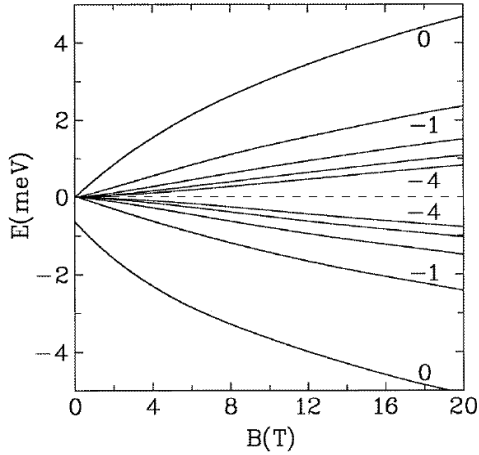

Figure 3. Binding energies of magneto-donor (negative) and repulsive energies of magneto-acceptor $A^{1-}$ (positive), related to the conduction LL $n = 0$. They are calculated for impurities at the interface ($z_0 = 0$) as functions of magnetic field for $N = 0$ and different $M \leq 0$. After Ref. [3].

One can describe possible magneto-optical transitions of the cyclotron-resonance type between MA states belonging to the $n = 0$ and $n = 1$ LLs, respectively. The selection rules for the left circular polarized radiation $\sigma_L$ are: $\Delta M = +1$ and $\Delta N = +1$. The corresponding optical transitions are indicated in Fig. 4. If both circular light polarizations are used in experiments the transition energies can be higher or lower than the free-electron cyclotron energy $\hbar\omega_c$. Intraband transitions of the CR-type between MA states were observed experimentally, see section 2.3 below.

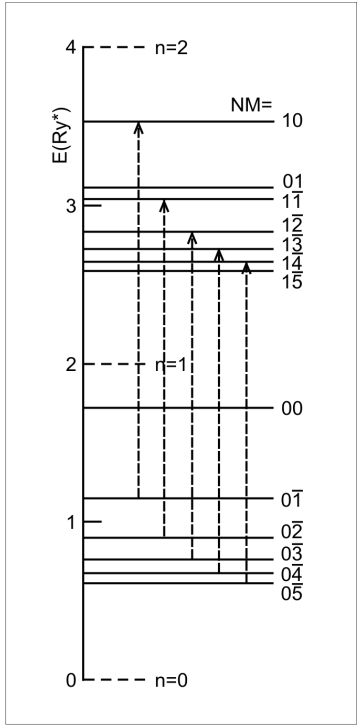

Figure 4. Magneto-optical transitions of CR-type between magneto-acceptor states in GaAs/GaAlAs heterostructure for $\sigma_L$ light polarization, obeying the selection rules $\Delta M = 1$, $\Delta N = 1$. The energies, calculated for $B = 10$ T, are plotted to scale in $Ry^* = 5.4$ meV. After M. Kubisa and W. Zawadzki (unpublished).

In conclusion of this theoretical section we quote the result of Bonifacie [9] on extension of the wave functions for magneto-acceptor, magneto-donor and free electron ground and first excited states $|N, M\rangle$, as calculated by the variational procedure for a magnetic field of B = 10 T. The lengths are given in Bohr radii for GaAs, i.e. in units of 98 Å. Figure 5 shows the 2D radial probability density $P(\rho) = (1/2\pi) \rho |\langle \rho, \varphi | N, M \rangle|^2$ for the three cases. It is seen that the extension of the MD states is the smallest and that of the MA states is the largest of the three. This agrees with the negative potential for MD and the positive one for MA.

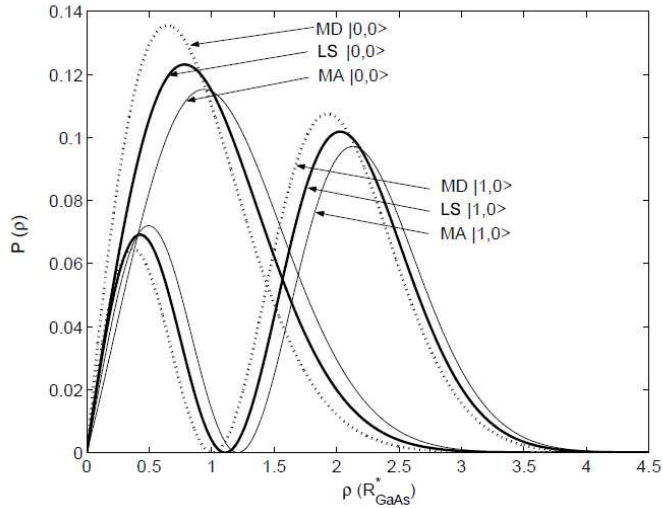

Figure 5. Calculated 2D radial probability density versus length (in effective Bohr radii for GaAs) of the ground and first excited electron states for magneto-acceptor (MA), magneto-donor (MD) and free electron Landau states (LS). Lengths are in Bohr radii for GaAs, i.e. in units of 98 Å. After Ref. [9].

## 2.2 INTERBAND PHOTO-MAGNETO-LUMINESCENCE

Photo-magneto-luminescence (PML) is a powerful method to study carrier excitations in semiconductors. There exist many experimental works on 2D structures using PML, see e.g. [10, 11]. The discrete magneto-acceptor states in the conduction band of GaAs/GaAlAs hetero-structures were studied experimentally with PML at the University of Montpellier. The first observation was that of Vicente et al. [12, 13] and the following description is based on this work. In PML experiments one illuminates the sample with light having energy much larger than the gap value. Excited electrons in the conduction band come quickly down to the conduction levels of high density of states and then recombine with the valence states emitting radiation which is detected. This way one explores simultaneously many conduction states. The electrons having high energies in the conduction band are relaxed mostly by optic phonons which can be emitted at low temperatures. The investigated single asymmetric quantum wells were donor-doped with Si atoms in the GaAlAs barrier on one side, see figure 6. Before the doping, the heterostructures were residual p-type. The residual acceptors were probably carbon atoms. After the donor doping the Fermi energy is in the conduction band, so that all the residual acceptors in the well are ionized.

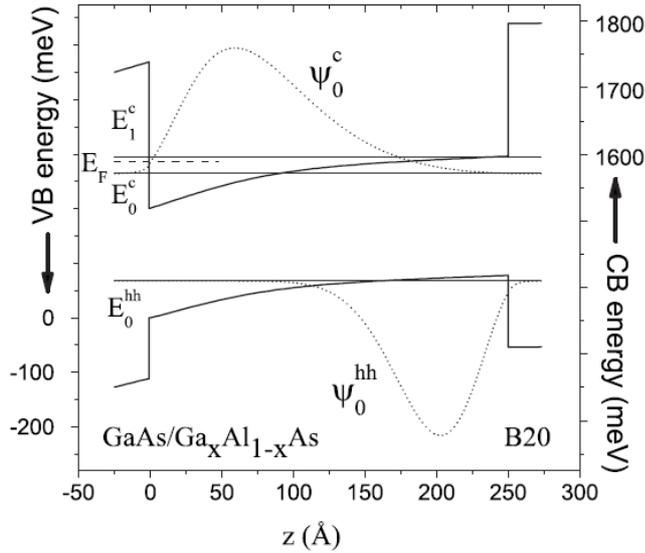

Figure 6. Subband structure of the asymmetric modulation-doped GaAs/$Ga_{0.67}Al_{0.33}As$ quantum well versus growth dimension, used for photo-magneto-luminescence experiments. The calculated conduction and heavy-hole subband energies and the corresponding wave functions are shown. The position of the Fermi level is indicated. After Ref. [14].

The fan chart of the observed PML energies is shown in Fig. 7. One sees a nonlinear behavior of energies as functions of magnetic field. This is typical for 2D structures and has been explained by electron density oscillations in the quantum well, see Ref. [14]. The solid lines without oscillations are drawn using heavy-hole LLs [15] and GaAs conduction LLs taking into account weak band's nonparabolicity [16, 17]. The strong transition A is identified as originating from the ground magneto-donor (MD) state in the conduction band and the so-called $1\beta$ state in the heavy-hole valence band. The binding MD energy (i.e. the difference between B and A lines) corresponds quite well to the theory for donors accumulated at the GaAs/GaAlAs interface ($z_0 = 0$), see Fig. 3. On the other hand, a striking behavior of the F transition in tilted magnetic fields is observed, as illustrated in Fig. 8. As the field forms an angle $\alpha$ with respect to the growth direction $z$, the peak F becomes weaker and by $\alpha = 30°$ it is not seen any more. This is in contrast with the behavior of the MD peak A which remains strong. The behavior of F transition characterizes the states of electrons localized by magneto-acceptors. In a tilted magnetic field the electron is less effectively confined by the joined effect of both fields. As a consequence, being repulsed by the ionized acceptor, the electron runs away in the x-y plane and the localized state disappears..

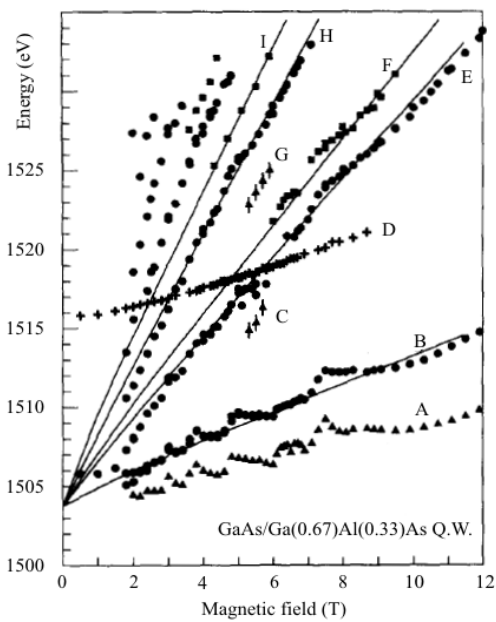

Figure 7. Energies of interband PML resonances observed on GaAs/GaAlAs heterostructure residually doped in the well with acceptors versus magnetic field parallel to the growth direction. The solid lines are theoretical. Transition F originates from the $A^{1-}$ centres, i.e. conduction electrons localized by ionized magneto-acceptors. After Ref. [12].

Figure 8. Photo-magneto-luminescence of GaAs/GaAlAs quantum well traced as a function of energy for external magnetic field ***B*** tilted at different angles $\alpha$ with respect to the growth direction. The projection of magnetic field on the $z$ direction $B_\perp = 8.1$ T is kept constant. As the angle $\alpha$ grows, transition F disappears while transition A remains almost constant. After Ref. [12].

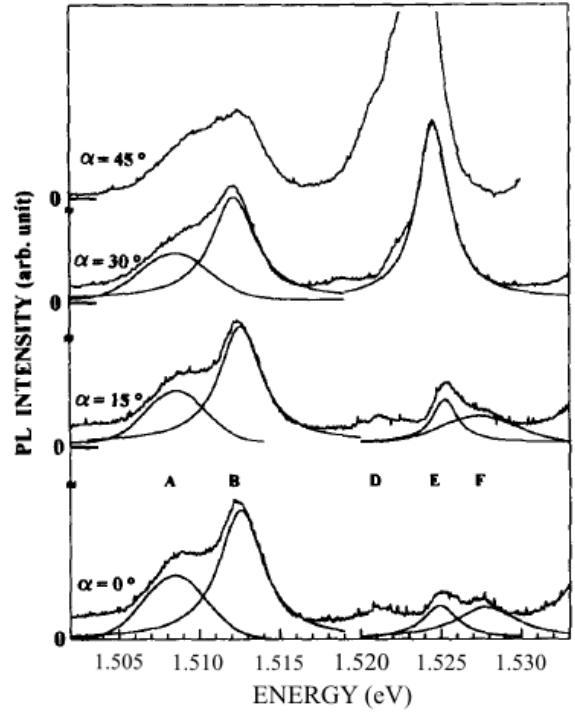

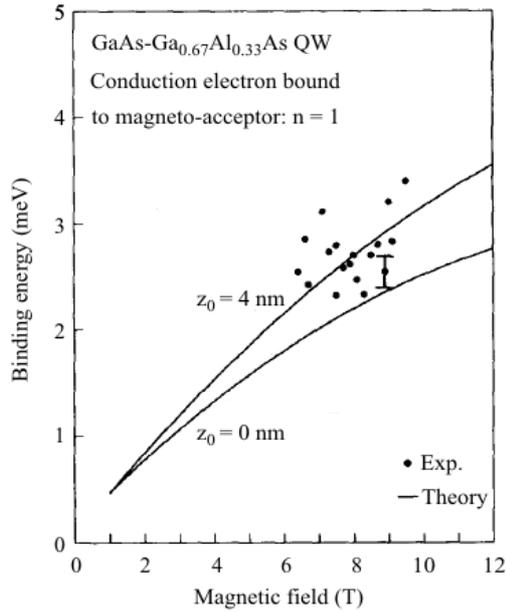

Figure 9. Repulsive energy of the conduction electrons localized by magneto-acceptors (associated with the $n = 1$ Landau level) versus magnetic field. Points – experimental data. Solid lines are theoretical, calculated after Ref. [3] (see section 2.1) for ionized acceptors located at the GaAs/GaAlAs interface ($z_0 = 0$ Å) or at the maximum of the envelope function ($z_0 = 40$ Å). After Ref. [12].

Figure 9 shows the experimental energy difference between the transitions F and E, which is interpreted as the repulsive energy of the conduction electron localized by the ionized MA. This energy is compared to theoretical results for the screened $A^{1-}$ centers located at the interface ($z_0 = 0$) or at the maximum of the envelope function in the quantum well ($z_0 = 40$ Å), as discussed above. The observed repulsive energies agree quite well with the theory presented in Section 2.1. As follows from Fig. 5, radius of the MA ground state is roughly 90 Å at $B = 10$ T so that, for a residual acceptor density of $10^{10}$ cm$^{-2}$, the acceptors can be safely treated as separate objects. The agreement shown in Fig. 9, together with the behavior in tilted magnetic field mentioned above, provides an experimental and theoretical evidence that one observes the states of conductions electrons localized by ionized magneto-acceptors. These are characterized by discrete repulsive energies above the corresponding Landau levels.

## 2. 3 INTRABAND MAGNETO-OPTICAL ABSORPTION

The discrete states of conduction electrons localized by ionized MA in GaAs/GaAlAs quantum wells were also observed using intraband magneto-optics, i.e. experiments of the cyclotron resonance type. In contrast to the interband photo-luminescence light emission described above, one investigates far-infrared light absorption in the vicinity of cyclotron frequency given by the energy difference between two consecutive Landau levels. This situation for the ground and first LLs is shown schematically in Fig. 4. The following account is based on the work of Raymond et al. [18] and Bonifacie et al. [19]. Three samples of GaAs/GaAlAs asymmetric quantum wells were investigated, δ-doped in the GaAs quantum well with Be atoms. This is a considerable technological improvement compared to the case described above in which the heterostructures contained residual acceptors (probably carbon atoms). The δ-doping means that the Be acceptors are put into the well at a well defined distance from the GaAlAs/GaAs interface. Such samples showed in the vicinity of cyclotron resonance not only the expected strong CR peak, but also rich spectra of small additional peaks.

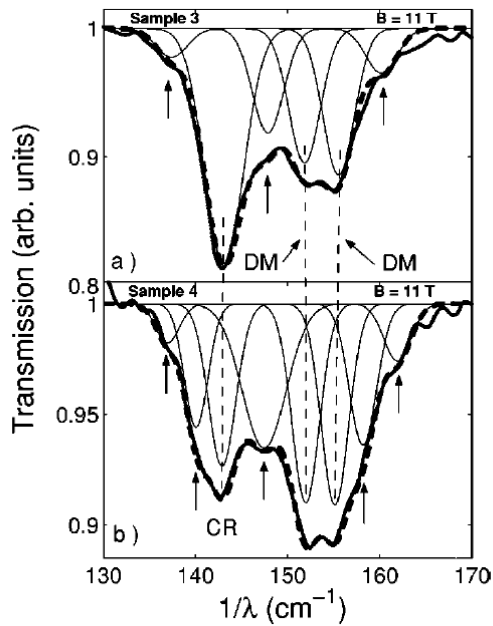

Figure 10. Infrared transmission spectra for two GaAs/GaAlAs QWs doped with Be acceptors at a magnetic field $B = 11$ T. In addition to the cyclotron resonance and two disorder modes (see description in section 4) one observes excitations evidenced by the deconvolution procedures. Solid lines – experiment, dashed lines – the sum of Gaussian peaks. (a) - sample 3, (b) – sample 4. After Ref. [19].

In order to determine the transition energies, deconvolutions of the spectra were carried out with the use of Gaussian peaks. Two examples of this procedure are shown in Fig. 10 for different samples. The experimental data are compared with the variational theory of Ref. [3] for magneto-acceptors, as shown in Fig. 11. The magneto-optical transitions are described by the selection rules for two circular light polarizations. According to these rules the transition energies can be larger or smaller than the CR energy. The smaller energies correspond to transitions from MA states above LL $n$ to the free electron state LL $n' = n + 1$, while the larger energies correspond to transitions from LL $n$ to MA states above LL $n' = n + 1$. The two maxima of transition energies, first around $B = 5.7$ T and second around $B = 10.8$ T, correspond most probably to the resonant screening of the acceptor potential, which is not included in the theory. (An analogous effect for magneto-donors was convincingly demonstrated both experimentally and theoretically in Ref. [20].) All in all, the above results give strong confirmation for the existence of discrete electron energies related to MA above the Landau levels. It should be mentioned that the two pronounced peaks at higher energies seen in Fig. 10 correspond to the so called "disorder modes", discussed in section 4.

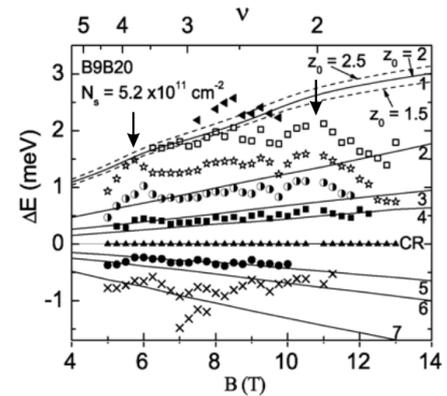

Figure 11. Energies of intraband magneto-optical transitions (counted from the CR energy) between the magneto-acceptor and Landau level states versus magnetic field. Solid lines show the calculated energies for $z_0 = 2$ nm from the interface. Vertical arrows indicate maxima of transition energies related probably to the resonant screening of acceptor potentials. After Ref. [19].

## 2.4 MAGNETO-TRANSPORT

Magneto-transport phenomena can supply an additional valuable information on the physics of acceptor-doped heterostructures. In experiments, one drives a current in the 2D plane along the $x$ direction and applies an external magnetic field $\boldsymbol{B}$ perpendicular to the 2D plane along the $z$ direction. This creates the transverse Hall field along the $y$ direction and changes electric conductivity in the $x$ direction. Measured values are usually expressed by resistance components $\rho_{xy}$ and $\rho_{xx}$. At low temperatures and higher magnetic fields the Landau quantization comes into play and one deals with the Quantum Hall Effect and the Shubnikov-de Hass effect. The transport effects supplement optical information in two ways. First, they depend strongly on the position of the Fermi energy which gives information on the thermal excitations of the localized states. Second, the electron localization by charged acceptors is directly reflected in the density of mobile electrons which is directly measurable by the electrical resistivities.

In this context we describe below the influence of discrete electron states in the conduction band related to charged magneto-acceptors on magneto-transport properties of GaAs/GaAlAs heterostructures. We will be concerned primarily with two effects: the so called "rain down" effect observed at weaker electric fields and the "boil off" effect observed at higher electric fields. The following account is based on the work of Raymond et al. [21], Bisotto et al. [22] and Kubisa et al. [23].

## 2. 4. 1 RAIN DOWN EFFECT

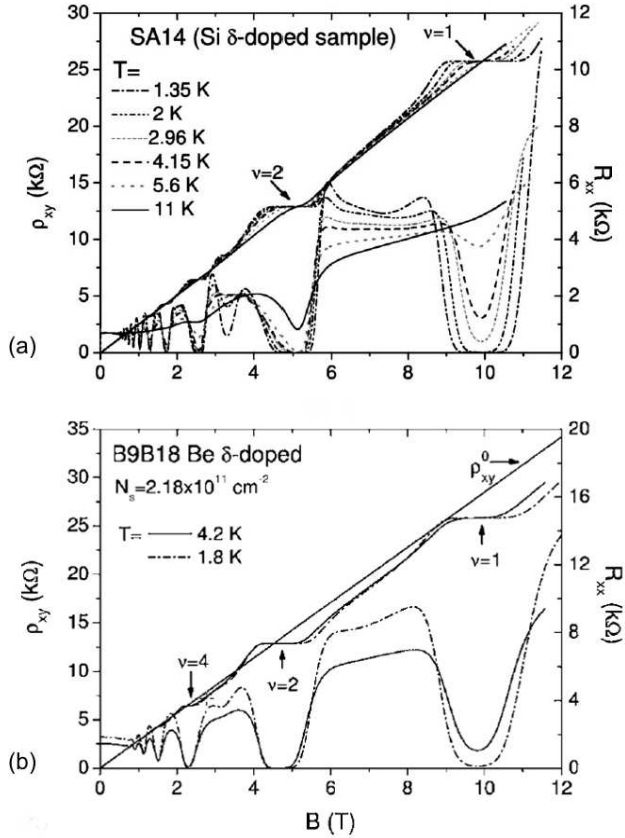

Figure 12. Resistivity components of GaAs/GaAlAs heterostructures at different temperatures versus magnetic field. (a) Sample doped additionally in the well with Si donors ($N_d = 2\times10^{10}$ cm$^{-2}$). (b) Sample doped additionally in the well with Be acceptors ($N_a = 1\times10^{10}$ cm$^{-2}$). After Ref. [21].

An important starting point in the analysis of magneto-transport at weak electric fields is the observation that quantum Hall lines measured on *all samples* at different temperatures exhibit common crossing points in the plateau regions. This property holds for samples additionally doped with acceptors, donors, as well as for the so called reference samples not additionally doped in the well. The meaning of "common crossing points" is illustrated in Fig. 12. It is seen that the common crossing points at six temperatures for the donor-doped sample (upper panel) coincide with the classical straight Hall line both for the asymmetric plateau at the filling factor $v = 2$, as well as for the symmetric plateau at $v = 1$. (The filling factor is defined in the standard way as $v = N_s(hc/eB)$). For the acceptor-doped sample (lower panel) there still exist common crossing points for two temperatures but the plateau corresponding to $v = 1$ is so much shifted to higher magnetic fields, that it falls completely below the straight Hall line. This striking result is common for *all* $v = 1$ plateaus in acceptor-doped samples, see Fig. 14 in this section and Figs. 19 and 23 in section 3. It is clear that the shift of $v = 1$ plateaus to higher magnetic fields goes beyond the plateau asymmetry observed for $v \geq 2$ plateaus on various samples and reviewed in section 3.

The analysis of various data indicates that it is the common crossing point of Quantum Hall (QH) lines at different temperatures that should be used to determine the electron density of conducting electrons at a given field $B$ according to the relation $N_s = B/e\rho_{xy}$. If this procedure is applied to the $\nu = 1$ plateau in Fig. 12b, one obtains $N_s = 2.3\times10^{11}$ cm$^{-2}$, i.e. a higher density than $N_s = 2.18\times10^{11}$ cm$^{-2}$ measured at low fields. The increase of $\Delta N_s = 1.2\times10^{10}$ cm$^{-2}$ compares quite well with the number of additional acceptors $N_a \cong 1\times10^{10}$ cm$^{-2}$ in this sample. The agreement between delta $\Delta N_s$ determined from the $\nu = 1$ plateau and number of additional acceptors was observed on other samples, as shown in Table 1. Similar feature was observed by Haug et al. [24], see Table 1 and section 3.

| Sample Ref. [21], [22] | Be acceptors $N_a$ ($\times 10^{10}$ cm$^{-2}$) | $z_0$ (Å) | $d$ (Å) | $N_s$ ($\times 10^{11}$ cm$^{-2}$) | $\mu$ ($\times 10^5$ cm$^2$/Vs) | $\Delta N_s$ ($\times 10^{10}$ cm$^{-2}$) |
|---|---|---|---|---|---|---|
| B9B18 | 1 | +20 | 250 | 2.18 | 0.8 | 1.2 |
| 35A52 | 0 | | 400 | 2.7 | 5 | |
| 35A53 | 2 | +25 | 400 | 2.5 | 0.88 | 2.1 |
| 35A54 | 2 / 2 | 0 / +25 | 400 | 2.25 | 0.36 | 3.8 |
| 35A55 | 4 | +25 | 400 | 1.36 | 0.53 | 3.5 |
| Sample Ref. [24] | 4 | +20 | 210 | 2.1 | 0.3 | 4 |

TABLE 1. Sample characteristics : $N_a$ – density of additional Be acceptors; $z_0$ - distance of Be $\delta$ layer from GaAs/GaAlAs interface; $d$ – spacer thickness; $N_s$ – density of 2D electrons; $\mu$ – mobility of 2D electrons at low temperatures; $\Delta N_s$ - additional electron density. In sample 35A54 there are two additional Be-doping layers. Reference sample 35A52 was not doped with acceptors. After Refs. [21], [22] and [24].

The above reasoning strongly suggests that the additional electrons "rain down" from the acceptor states in the conduction band to the corresponding Landau levels $0^+$ and $0^-$ once the Fermi energy passes below the MA levels with the increasing magnetic field, see note [25]. The process of rain down is schematically illustrated in Fig. 13. In GaAs under usual conditions the spin $g$-value is very small and the spin splitting is considerably smaller than the acceptor binding energy. It means that the $0^-$ LL is also affected by the increasing density of conduction electrons. When the Fermi energy passes between the $0^-$ and $0^+$ LLs, the exchange enhancement of the spin $g$-value takes place and the spin splitting strongly increases [26]. This is the range of $v = 1$ filling factor in which the spin splitting can become larger than the MA binding energy. But at such high fields all electrons are already in the lowest $0^+$ LL, which means that the rain down is completed and the density of conducting electrons has increased.

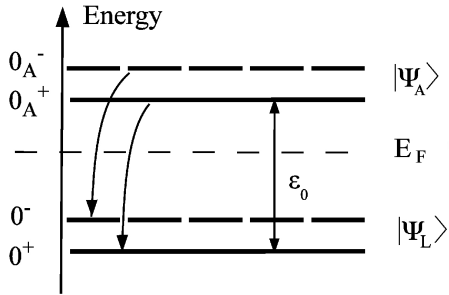

Figure 13. Free Landau levels and discrete MA levels in the Quantum Hall Effect regime at high magnetic fields (schematically). As the magnetic field is increased and the Fermi energy passes below MA levels, the localized acceptor electrons rain down into delocalized Landau states, see text for details. After Ref. [22].

Figure 14. Experimental Hall resistivities $\rho_{xy}$ at low currents and $T = 1.5$ K, vs. magnetic field normalized by the value of $B_2$, corresponding to the filling factor $v = 2$, for four samples of different acceptor densities (see Table 1). The shift $\Delta B$ of the $v = 1$ plateau increases with the acceptor density, which is a manifestation of the rain down effect. After Refs. [21, 22].

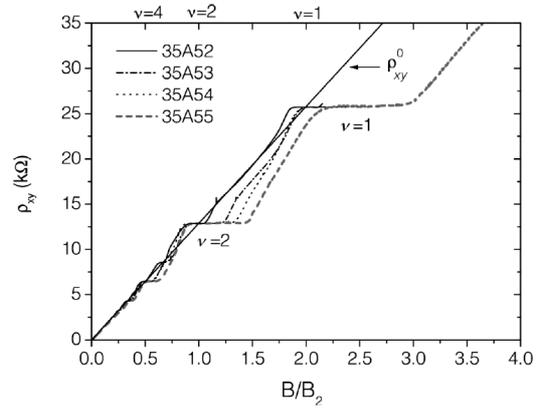

Figure 14 shows the quantum Hall results of four samples containing different densities of additional acceptors. In order to be able to compare directly different samples it is necessary to normalize magnetic field on the abscissa to the value of $B_2$ corresponding to the filling factor $v = 2$ for each sample. It is seen that with the increasing content of additional acceptors the quantum Hall lines shift more and more to higher fields in the range of $v \geq 2$. This corresponds to the rain down effect, as indicated above. As follows from Fig. 14, this situation occurs at the onset of $v = 2$ plateau.

One could possibly suppose that the increase of $N_s$, beginning near $v = 2$ and continuing all the way through $v \leq 1$, is due not to the rain down effect from the acceptor levels but rather to the transfer of electrons from the reservoir to the well, as proposed by Baraff and Tsui [27] for GaAs/GaAlAs heterostructures and reviewed in some detail by the present authors [28]. However, as follows from the analysis of the electron transfer between reservoir and quantum well, the latter does not lead to a constant shift of quantum Hall lines with respect to the classical Hall straight line, but to oscillations of the quantum lines around the straight line. The reason is that, in the reservoir mechanism, the free electron density oscillates as a function of $B$ around the initial density at $B = 0$.

## 2. 4. 2 BOIL OFF EFFECT. RESULTS AND INTERPRETATION

Finally, we consider the discrete states of conduction electrons bound to acceptors in GaAs/GaAlAs heterostructures manifested in magneto-transport phenomena at high electric fields. It is known that at low electric currents one observes quantum effects in $\rho_{xy}$ and $\rho_{xx}$ magneto-resistivities. At higher driving currents in standard samples there occurs the well known breakdown of Quantum Hall Effect. In contrast, for acceptor-doped samples, and particularly for sample 35A55, whose density of Be atoms $N_a = 4\times10^{10}$ cm$^{-2}$ is nearly equal to one third of the electron density $N_s$, the behavior is completely different.

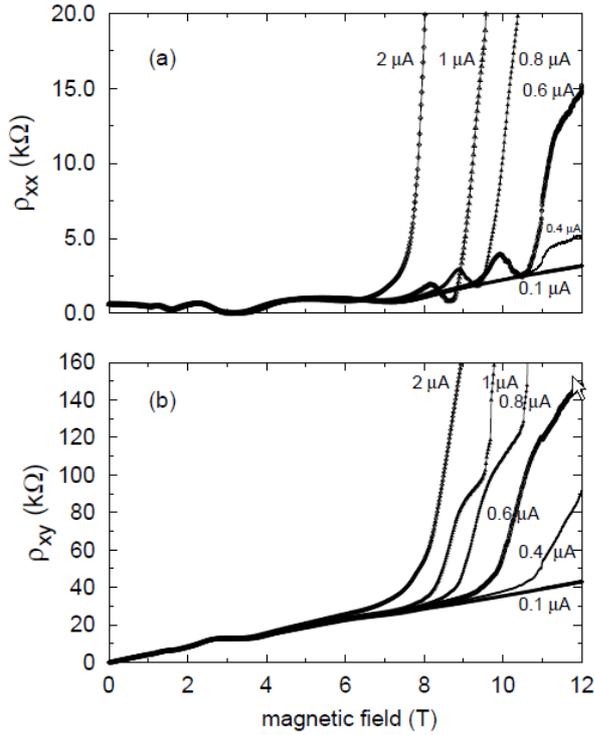

Figure 15. Transport characteristics of acceptor-doped GaAs/GaAlAs heterostructure, measured versus magnetic field for different stabilized currents in dc experiments. At higher magnetic fields (resulting in higher Hall fields) a strong increase of $\rho_{xx}$ and $\rho_{xy}$ resistivities is observed indicating a decrease of conducting electron density. After Ref. [22].

Experiments carried out for different stabilized currents are shown in Fig. 15. It is seen that, at higher currents, both resistivities $\rho_{xx}$ and $\rho_{xy}$ dramatically increase. This increase is interpreted as a consequence of the strong decrease of mobile electron density $N_s$. In experiments shown in Fig. 15 the Hall electric field $F_y$ is always stronger than the driving field $F_x$. If the driving current $J_x$ is stabilized, a higher magnetic field results in a higher Hall field $F_y$. If this field diminishes the density of mobile electrons, both resistivities increase and the constant current $J_x$ requires a higher driving field $F_x$. Thus the process has an avalanche character which is seen in Fig. 15. As we demonstrate below, a sufficiently strong Hall field $F_y$ induces transitions of Landau electrons to the localized acceptor states, which decreases the density of conducting electrons. This process has been called the "boil off effect", in contrast to the well known "freeze-out effect" for electrons in magneto-donors, see Ref. [29].

The boil off effect is also manifested in experiments using stabilized electric voltage rather than currents. In this case, instead of applying a dc current, an ac voltage is applied on opposite ends of the sample. For samples without additional acceptors, when the voltage is increased, the standard breakdown of Quantum Hall Effect is observed. On the other hand, in the acceptor doped sample an increase of $\rho_{xx}$ and $\rho_{xy}$ at higher voltages is observed, see Ref. [22]. These results can also be explained by the decrease of the mobile electron density $N_s$ due to the boil off effect, but this time the process does not have an avalanche character.

We do not go here into theoretical description of electron transitions induced by impurities or phonons, this can be found in Ref. [22]. Instead, we show schematically that, in the presence of a sufficiently high Hall electric field, electrons can transfer between free Landau states and MA states $A^{n-}$ in the boil off process. This increases the occupation of localized MA states and diminishes the density of conducting electrons. In the Quantum Hall Effect regime an electron is subjected to an external magnetic field $\boldsymbol{B}$ (in the $z$ direction) and the Hall transverse electric field (say, in the $y$ direction). Thus one deals with the electron in crossed electric and magnetic fields. It is convenient to choose the vector potential for $\boldsymbol{B}$ using the gauge $\boldsymbol{A} = (-By, 0, 0)$. Then the quantum eigenenergies for the problem are

$$E = \hbar\omega_c\left(n+\frac{1}{2}\right) + eF_y y_0 - \frac{1}{2}m^*c^2\frac{F_y^2}{B^2}, \qquad (7)$$

where $y_0 = k_x(\hbar c/eB)$ is the center of magnetic orbit. If we neglect the shift of all levels due to the last term in Eq. (7), the energies as functions of $y_0$ and the resulting density of states can be represented schematically in Fig. 16. It is seen that the transfer processes can take place when the Landau level at the right sample edge and the $A^{1-}$ level at the left sample edge begin to overlap.

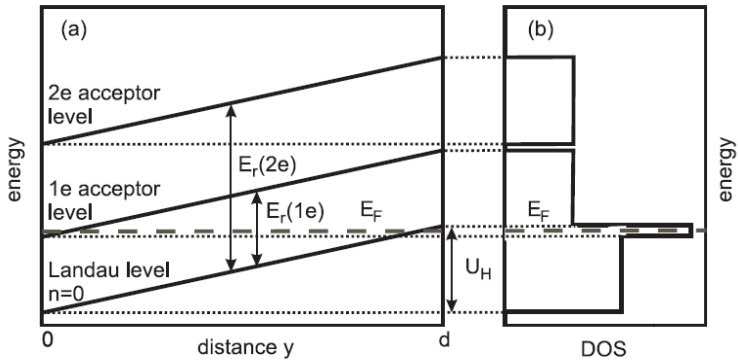

Figure 16. (a) Free electron and electron repulsive energies in the sample in crossed electric and magnetic fields. (b) Corresponding density of states (DOS). The boil off process takes place when the Landau level at the right sample edge and the $A^{1-}$ level at the left sample edge begin to overlap. After Ref. [23].

The experimental resistivities $\rho_{xx}$ and $\rho_{xy}$, shown in Fig. 15, can be used to determine the density of conducting electrons with the help of relation

$$N_s = \frac{B\rho_{xy}}{e\left(\rho_{xy}^2 + \rho_{xx}^2\right)}. \qquad (8)$$

The obtained results, plotted as functions of *B* for different currents, are shown in Fig. 17. At lower fields *B* one can see two linear increases of $N_s$, which can be attributed to the electron transfer from a reservoir, see Ref. [28]. In the range of 4 T ≤ *B* ≤ 7 T the density $N_s$ is higher than the initial value $N_s(0)$, which one can attribute to the rain down effect. Finally, at fields *B* ≥ 7 T one observes a dramatic decrease of $N_s$ which, according to our analysis, is due to the boil off phenomenon. At higher magnetic fields, which result in higher Hall electric fields, the mobile electron density $N_s$ drops to almost zero. This means that the relatively small number of acceptors in the well ($N_a = 4 \times 10^{10}$ cm$^{-2}$) can localize practically all free electrons. In other words, each acceptor localizes up to four electrons. It is argued below that this is indeed possible since the confinement of electron motion around an ionized acceptor is not due to the Coulomb attraction.

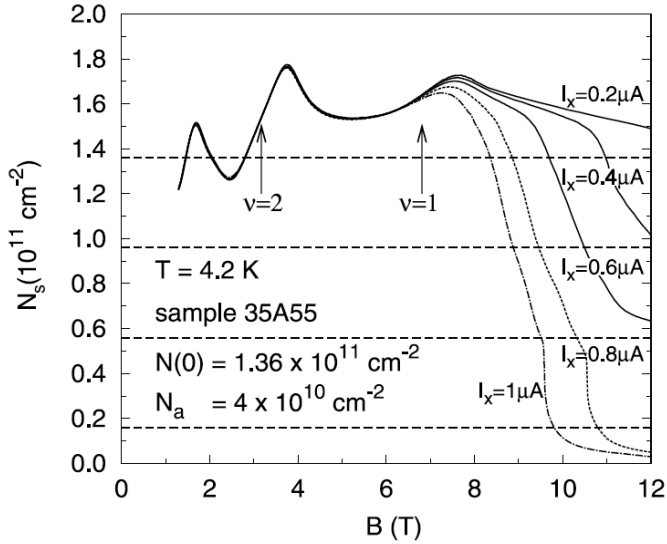

Figure 17. Conducting 2D electron density $N_s$ for different currents versus magnetic field, as determined experimentally from the resistivities $\rho_{xx}$ and $\rho_{xy}$ shown in Fig. 15. At higher magnetic fields $N_s$ drops to almost zero as a result of the boil off transitions from the Landau states to the localized MA states $A^{n-}$. After Ref. [23].

### 2. 4. 3 THEORY FOR TWO LOCALIZED ELECTRONS

It was demonstrated in the previous section that a negatively charged acceptor in the presence of a magnetic field can localize more than one conduction electron. This experimental result was confirmed by the theory of a negatively charged acceptor ion localizing two conduction electrons in a GaAs/GaAlAs quantum well worked out by Kubisa et al. [23]. Below we outline main features of this treatment. It generalizes the theory for one localized electron reviewed in section 2. It also goes beyond the one-electron theories of Refs. [4, 5]. One considers a pair of conduction electrons at positions $r_j = (x_j, y_j, z_j) = (\rho_j, z_j)$, where *j* = 1, 2, in a heterojunction described by the potential *U*(*z*). The electrons move in the presence of a magnetic field ***B*** || *z* and interact with an ionized acceptor located at $r_0 = (0, 0, z_0)$. The initial Hamiltonian for the problem reads

$$H = \sum_{j=1,2}\left[\frac{1}{2m^*}\left(p_j + \frac{e}{c}A_j\right)^2 + V(|r_j - r_0|) + U(z_j)\right] + V(|r_1 - r_2|), \qquad (9)$$

where $V(r)$ represents the Coulomb potential, and $A_j = \left(-\tfrac{1}{2}By_j, \tfrac{1}{2}Bx_j, 0\right)$ is the vector potential of the magnetic field. Using the center of mass coordinate $R$ and the relative coordinate $r$ of the electron pair in the x-y plane one reduces the 3D problem to an effective 2D problem similarly to the one-electron system described above [3, 30]. Energies of the resulting 2D problem are calculated variationally using the two-electron function

$$F_{2e}(r,R) = \frac{1}{2\pi\alpha\beta}\exp\left(-\frac{R^2}{2\alpha^2} - \frac{r^2}{8\beta^2}\right). \qquad (10)$$

The variational parameters $\alpha$ and $\beta$ have simple physical meanings: $\alpha$ is the radius of the center-of-mass motion around the acceptor, while $\beta$ is the average distance between two electrons. Since the electrons are indistinguishable, the total wave function must change sign under the permutations of particles. To satisfy this condition, the ground state (10) corresponds to the singlet state having electron spins in opposite directions. In order to compare the two-electron energies with the one-electron ones, as discussed in section 2.1, the ground state energy is also calculated for the one-electron state localized by a negatively charged acceptor, with the use of trial function

$$F_{1e}(\rho) = \frac{1}{\lambda\sqrt{2\pi}}\exp\left(-\frac{\rho^2}{4\lambda^2}\right), \qquad (11)$$

where $\lambda$ is a single variational parameter. This function is simpler than that used in Section 2.1.

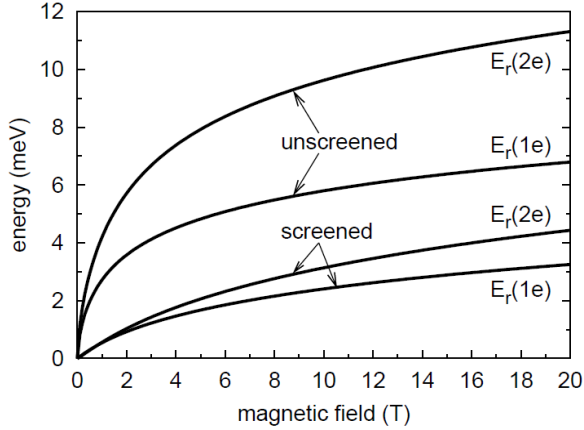

Figure 18. Calculated one-electron and second-electron repulsive energies of conduction electrons confined to the proximity of a negatively charged acceptor ion by a combined effect of the quantum well and a magnetic field in a GaAs/GaAlAs heterostructure. Here $E_r(1e)$ is the energy required to localize the first electron on a negative acceptor, and $E_r(2e)$ is the energy required to add the second electron to the acceptor already occupied by one electron. After Ref. [23].

Figure 18 shows the main results of the theory for a GaAs/GaAlAs heterostructure. It is seen that both one-electron and two-electron energies vanish at $B = 0$, which reflects the fact that at a vanishing magnetic field there are no localized electron states. One can also see that the screening considerably diminishes the repulsive energies. The repulsive one-electron energies are lower than those presented in section 2.1 because of the simpler variational function, as mentioned above. All in all, it follows from Fig. 18 that the energy necessary to add the second conduction electron to an acceptor ion is higher than that necessary for the first electron. As a consequence, one expects that, in an experiment, one first populates all available acceptor ions with single electrons and only then begins to populate them with second electrons. It is seen from Fig. 17 that this is what indeed happens experimentally. An indication that one deals with the localization of consecutive electrons by acceptor ions is that one observes accidents, at least on some of the $N(B)$ curves, in intervals equal to the number of acceptors in the well, see Fig. 17.

It should be mentioned that the increase of conducting electron density, which we attribute to the rain down effect, was observed also in experiments of other authors, see below. In addition, Buth et al. [31] in their magneto-transport experiments found at higher magnetic fields a strong decrease of conduction electron density $N_s$, which the authors ascribed to "localization of the electrons into the droplet phase". This observation is similar to ours, but we interpret it above as the localization by acceptor ions. The localized conduction states due to repulsive potentials in quantum wires were described by Gudmundsson et al. [32]. Instead of "repulsive energies" the authors used the term "negative binding energies".

## 3. ASYMMETRY EFFECTS IN MAGNETO-TRANSPORT

Here we consider the second phenomenon caused by the presence of acceptor impurities in quantum wells, i.e. an asymmetry of the Quantum Hall plateaus and of the Shubnikov–de Haas (ShdH) minima in magneto-transport. Interpretation of the asymmetry goes beyond the one-acceptor picture presented above.

### 3.1 EARLY EXPERIMENTS

To our knowledge, the effect of asymmetry was first reported by Furneaux and Reinecke [33] on Si metal-oxide-semiconductor field-effect transistors in which driftable $Na^+$ ions in the oxide, acting as acceptors, influenced the width and position of the QH plateaus. The authors interpreted their results in terms of an asymmetric distribution of the localized states in the tails of Landau levels. Haug et al. [24, 34] observed a similar effect in GaAs\GaAlAs heterojunctions: in a sample doped in the well with Si donors the QH plateaus were slightly asymmetric toward lower magnetic fields, while a sample doped with Be acceptors showed the corresponding plateaus more strongly shifted to higher magnetic fields. Their results for the acceptor-doped sample are shown in Fig. 19. It can be seen that the highest QH plateau for $v = 1$ does not cross the line of the "classical" Hall resistance $\rho_{xy} = B/eN_s$. This result is typical for the acceptor-doped samples, see the discussion in section 2.3.1. It was suggested that the asymmetries of QH plateaus were due to the corresponding changes in the density of states (DOS) for the donor-doped and acceptor-doped samples. However, a theory based on the so called self-consistent T-matrix approximation, while showing clear differences between donor-doped and acceptor-doped samples, did not give good overall description of the observed data.

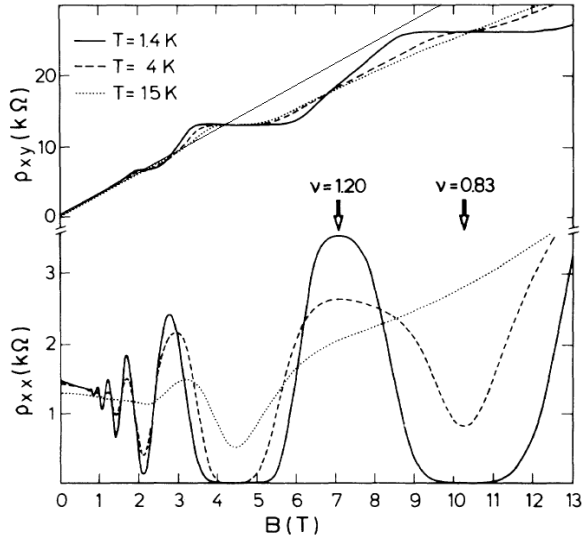

Figure 19. Resistivities $\rho_{xx}$ and $\rho_{xy}$ versus magnetic field measured on GaAs/GaAlAs heterostructure doped in the well with Be impurities. The thin straight line indicates $\rho_{xy} = B/eN_s$ with $N_s$ determined at low $B$ values. $N_s = 2.1\times10^{11}\text{cm}^{-2}$, $N_a = 4\times10^{10}\text{ cm}^{-2}$. After Ref. [24], see also Table 1.

### 3.2 THEORY AND INTERPRETATION

Further progress on the asymmetry of the QH plateaus and ShdH minima in quantum magneto-transport was reported in Refs. [35] and [36]. Our description of the subject is based mostly on this work. As we indicate below, the asymmetry is also partly related to the discrete MA states above the Landau levels, described in Section 2.

First calculations of the density of states for 2D electrons in a magnetic field were carried out by Ando et al. [7, 37, 38] using the self-consistent Born approximation (SCBA) and treating the perturbative potential as delta functions (short range potential) without taking into account the LL mixing. This resulted in symmetric semi-elliptic DOS. The LL mixing and the electron screening were introduced in Refs. [39, 40]. The resulting DOS showed a smoother and more realistic shape than that resulting from short range impurities. Further studies included acoustic phonon and electron-electron scattering. The calculations, however, were carried by including only a few terms in the perturbation series (SCBA) which always yielded equally spaced LLs and symmetric DOS for each LL. This was not enough to describe experimental results of Refs. [24] and [34].

A proper theory should consider a realistic situation in which a GaAs/GaAlAs heterostructure is first modulation-doped with Si donors in the GaAlAs barrier to provide 2D electrons in the well. The ionized Si donors create a smooth attractive potential for the 2D electrons. Then the structure is doped inside the GaAs well with either Si donors or Be acceptors which provide sharp attractive or repulsive potentials, respectively. As a first step in the theory, Bonifacie et al. [36] represented the donor and acceptor potentials in the GaAs well by Gaussian functions with two adjustable parameters and showed that one could reproduce the MD and MA energies calculated variationally in Ref. [3].

Then the single particle Green`s function was calculated, in which the selfenergy was computed within the so-called fifth Klauder's approximation [41]. Finally, the DOS was related to the averaged Green's function.

The DOS for the lowest LL in acceptor-doped structures calculated at the high field limit of 10 T is shown in Fig. 20. (Similar results for the donor-doped structures were reported by Ando [38]). One observes a formation of impurity bands above the LL at some critical acceptor densities. Each isolated band, which corresponds to levels of different angular momentum ($m = 0, 1, 2, 3, \ldots$), is exactly centered on the binding energy calculated in Ref. [3]. As the acceptor concentration increases, the impurity bandwidth increases as well, until overlapping at the critical density $N_a = 2 \times 10^{10}$ cm$^{-2}$ at $B$ = 10 T. More generally, the critical density depends on the position of the doping layer, the 2D electron density $N_s$ and the field $B$. For sufficiently high magnetic fields the impurity band is separated from the corresponding Landau level. For intermediate and low magnetic fields one must take into account the mixing between Landau levels due to the impurity potentials. Higher LLs are less perturbed by the impurities than the lower ones which results in an unequal level spacing. This effect is sometimes called "unharmonicity", see Ref. [36]. For low LLs the maxima of DOS are not shifted much as compared to the unperturbed levels, but the additional DOS on the higher-energy sides plays an important role in the plateau asymmetry. For donor-doped samples the interaction with donor ions is attractive, the additional DOS appears on the lower-energy sides of LLs and the plateau asymmetry is opposite.

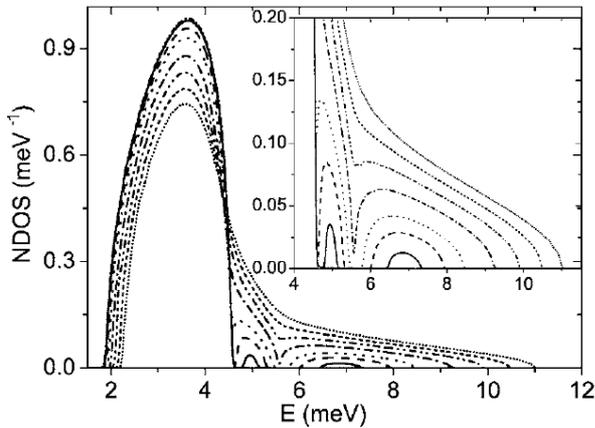

Figure 20. Normalized densities of states for $n = 0$ Landau level at $B = 10$ T versus energy, calculated for different acceptor densities. GaAs/GaAlAs heterojunction is δ-doped with two layers: the first has $N_d = 4 \times 10^{11}$ cm$^{-2}$ of Si atoms at $z_0 = 500$ Å, the second contains Be atoms at $z_0 = 0$. Different curves correspond to different density of the second layer: $N_a = 0$; 0.1; 0.5; 1; 2; 3; 4; $5 \times 10^{10}$ cm$^{-2}$. Inset shows details of the high-energy region. After Ref. [36].

Calculating the Hall resistivity $R_H$ one can compare the theory with experiment. It is assumed for simplicity that only one state per LL is delocalized and its energy $E_n$ is given by the maximum of DOS. In addition, it is assumed that each delocalized state contributes $e^2/h$ to the Hall conductivity. Within this simple approach, the adiabatic conductivity is given by

$$R_H^{-1} = \frac{e^2}{h} \sum_n f(E_F - E_n), \qquad (12)$$

where $f$ is the Fermi distribution and $E_F$ is the Fermi energy. As seen from Eq. (12), the Hall resistance depends on the relative positions of the Landau energies $E_n$ and the Fermi energy $E_F$. The behavior of $E_F$ is governed by the condition $N_s = \int_{-\infty}^{\infty} D(E) f(E_F - E) dE$, where $D(E)$ is the DOS. For given $N_s$ and magnetic field one calculates the integral and determines $E_F$. It turns out that the behavior of $E_F$ is markedly influenced by the additional DOS above LL, as seen in Fig. 20. It is this feature that is responsible for the asymmetry of the QH plateaus [9].

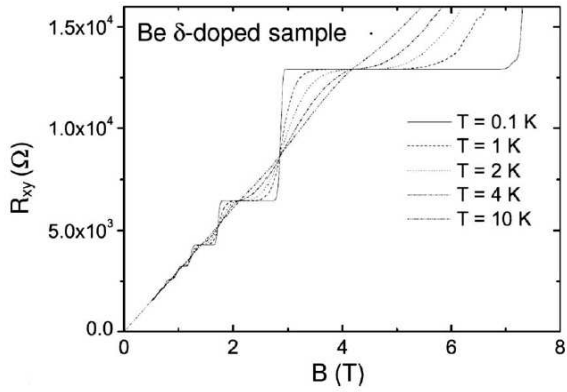

Figure 21. Theoretical Hall resistance of GaAs/GaAlAs heterostructure at different temperatures versus magnetic field for sample doped with Be acceptors at the interface ($N_a = 2 \times 10^{10}$ cm$^{-2}$). The calculation is based on asymmetric DOS shown in Fig. 20. The asymmetry of QH plateau is clearly seen. After Ref. [21].

Figure 21 shows theoretical results for the Hall resistance at different temperatures in an acceptor-doped sample, as calculated with the above procedure. The asymmetry of the Hall plateaus is quite pronounced, it goes in opposite directions for donors and acceptors and it increases rapidly when the temperature decreases. The important feature is that, for different $T$, all calculated Hall lines cross at the same horizontal inflection point. This property is common for all calculations related to the density of states, equally well for donor-doped, acceptor-doped and undoped samples.

## 3.3 LATER RESULTS

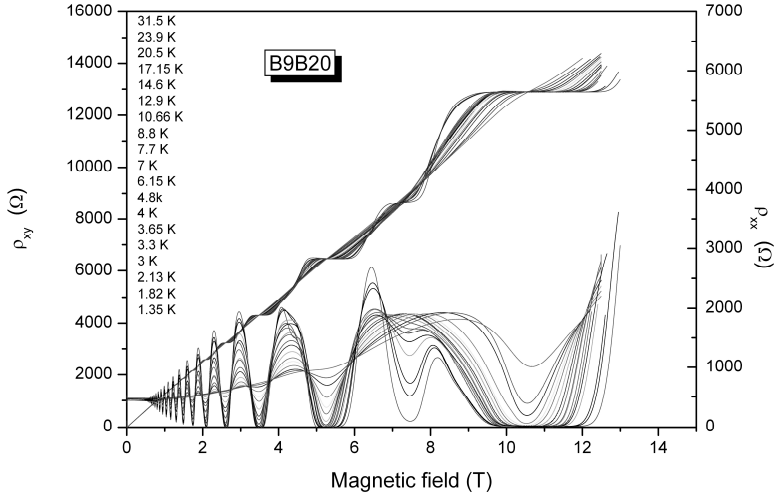

Figure 22. Experimental magneto-transport data taken at various temperatures on GaAs/GaAlAs heterostructure ($N_s = 5.0 \times 10^{11}$ cm$^{-2}$) additionally doped with Be acceptors in the well 20 Å from the interface ($N_a = 2 \times 10^{10}$ cm$^{-2}$). Asymmetry of the QH plateaus at $v$ = 2 and 4 and common crossing points of various QH lines are clearly visible. After S. Bonifacie and A. Raymond (unpublished).

Figure 22 shows experimental magneto-transport data taken at different temperatures on an acceptor-doped GaAs/GaAlAs heterostructure. It can be seen that the quantum Hall plateaus exhibit asymmetry with respect to the classical Hall straight line and that the quantum Hall curves for various temperatures cross at the common points. Thus the theoretical predictions based on the asymmetric DOS, as presented in Fig. 21, are fully confirmed experimentally. However, the above agreement based on asymmetric DOS applies to the filling factors $v \geq 2$. For the filling factor $v = 1$ the shift of the QH plateau requires additional explanations.

As observed in Ref. [24] and then confirmed in other investigations, this plateau is not really asymmetric but simply *shifted* to higher magnetic fields. In fact, the plateau is symmetric but with respect to the shifted crossing point, common to QH lines measured at various temperatures. This shift is so large that often the whole plateau is below the straight classical Hall line, see Figs. 14 and 19. We already explained in some detail in section 2. 3. 2 that, according to our understanding, the shift of whole $v = 1$ plateau is due to the "rain down" of electrons from the localized acceptor $A^{1-}$ states to the delocalized Landau states. As a consequence, the density of conducting electrons increases so that the center of the shifted plateau corresponds to a higher density $N_s$. Two features corroborate this interpretation. First, the Hall lines measured at different temperatures have again a common crossing point in the middle of the shifted plateau. Second, the shifted center of the plateau for $v = 1$ corresponds to the "new" electron density $N_s$' such that the electron density increase is equal to the number of acceptors $N_a$. Thus, in the data shown in Fig. 19, the center of shifted plateau corresponds to the "new" density $N_s' = 2.5 \times 10^{11}$ cm$^{-2}$. The additional density is $\Delta N_s = 4 \times 10^{10}$ cm$^{-2}$, which corresponds to the number of acceptors. Similar agreement was found in the data of Ref. [22], as quoted in Table 1.

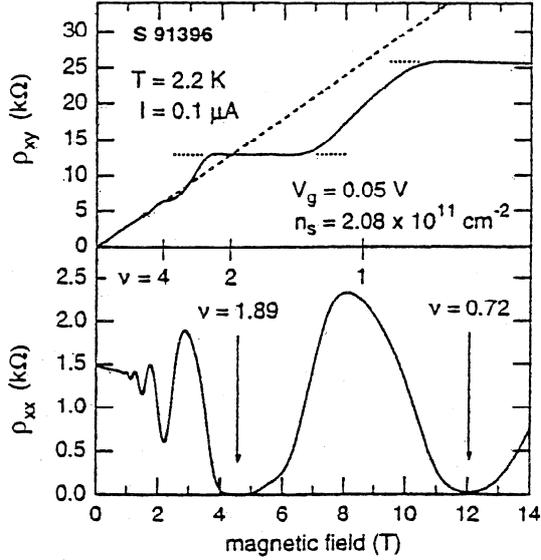

Figure 23. Plateaus of the Hall resistance and minima of the longitudinal resistivity, measured on GaAs/GaAlAs heterojunction doped with Be acceptors. Both plateaus and minima are shifted to higher magnetic fields. After Ref. [31].

To conclude this section we quote in Fig. 23 similar results of Buth et al. [31], in which the QH plateau corresponding to $v = 1$ is shifted to higher magnetic fields beyond the straight classical Hall line. The authors mention a possibility of an increase of electron density $N_s$ by 40% in order to explain their data. Indeed, looking at Fig. 23 it is difficult to believe that the highest plateau corresponds to the unchanged electron density. In our interpretation, the suggested increase of $N_s$ is a result of the electron rain down from the acceptor levels.

## 4. DISORDER MODES

Finally, we review magneto-optical properties of the so called "disorder modes", characteristic of the 2D systems doped with acceptors. The term "disorder mode" (DM) was introduced in Ref. [19]. Similar modes were seen first in Si-MOS devices, but later more thoroughly investigated in GaAs/GaAlAs heterostructures.

### 4. 1 EARLY STUDIES

Mott [42] and Stern [43] suggested that, at low surface electron densities, because of potential fluctuations at the Si surface there may occur localization of electronic states. Such potential fluctuations can arise from immobile charges in the oxide or at the interface. Transport experiments of various authors were interpreted in terms of the expected electron localization for sufficiently small electron densities ($N_s < 10^{12}$ cm$^{-2}$), see for example Refs. [44, 45]. Kotthaus et al. [46] first analyzed corresponding anomalies in the cyclotron resonance. It was observed that the resonant magnetic field at a fixed frequency depended on the 2D electron density $N_s$ and it was suggested that the localized electrons were represented by a set of harmonic oscillator levels of the frequency $\omega_{res}$, where $\omega_{res}^2 = \omega_0^2 + \omega_c^2$, with $\omega_c$ being the unperturbed CR frequency. This expression followed from an assumption that the localization was due to a 1D parabolic potential characterized by the frequency $\omega_0$. The electrons were assumed to occupy the surface in little puddles where the electrostatic potential has a local minimum. The above formula was subsequently used in many descriptions of experimental data and $\omega_0$ was often called the "offset". The increased splitting of the Landau levels due to a binding potential is a very general result. In an early theoretical description of localized electron states in a magnetic field Mikeska and Schmidt [47] assumed existence of 2D harmonic oscillator wells and averaged over four different well sets. In their model the effective CR frequency increases with increasing 2D electron density $N_s$, in agreement with the observations [46]. The model of localization represented by a harmonic oscillator in a magnetic field was considered in some detail by Wilson et al. [48]. The main shortcoming of this approach is that it predicts two disorder modes whereas one observes only one higher-frequency mode.

The investigation of GaAs/GaAlAs heterostructures doped with acceptors was initiated at Max Planck Institute in Stuttgart [49-52]. First samples were irradiated by electrons at low temperatures which introduced defects, mainly vacancy-interstitial pairs in the As sublattice. The defects act as traps for electrons and negatively charged occupied traps are repulsive centers (acceptors) for 2DEG. One can partially or completely anneal the traps at higher temperatures, thus changing the density of acceptors. The cyclotron resonance and quantum transport were studied on irradiated heterostructures and it turned out that the presence of negative defects influenced the effective cyclotron frequency $\omega$. The experimental results are quoted in Fig. 24.

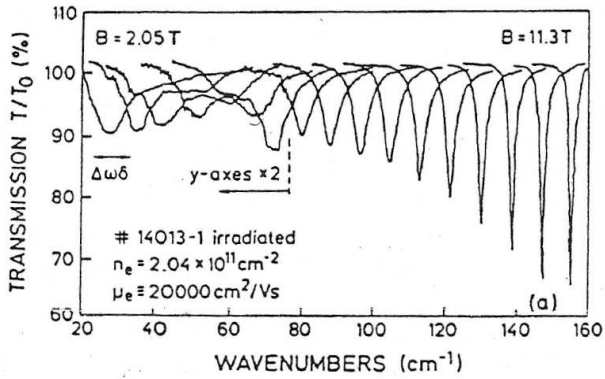

Figure 24. Cyclotron resonance transmission spectra taken on an irradiated GaAs/GaAlAs heterostructure at $T = 2.2$ K and constant magnetic fields of equal increment. At $B > 4$ T one resonance appears, at $B < 4$ T two resonances are observed: the cyclotron resonance and disorder mode. At high fields the DM line becomes narrow. After Ref. [50].

The following formula for the DM frequency was deduced:

$$\omega_{ef}^2 = \omega_s^2 + \omega_c^2, \qquad (13)$$

where $\omega_s^2$ is offset squared and $\omega_c = eB/m^*c$ is the standard CR frequency. The results, plotted in the form: $\omega^2$ versus $B^2$, are shown in Fig. 25. It is seen in the inset that the offset decreases with decreasing number of acceptors. For the reference sample (no acceptors) no offset is observed. For $B <$ 4 T one observes two resonances at each $B$. This is unexpected since the irradiation introduces acceptor-like centers in the whole volume of heterostructure. Thus the whole 2D channel should be homogenously doped and produce one, even if modified, frequency. The fact that the offset depends on the acceptor density means that the effective CR frequency is governed by a collective property of acceptor distribution. The shift of the disorder mode to higher frequencies resembles the behavior of Si-MOS systems. It is seen in Fig. 24 that at high fields the DM resonance becomes very narrow. This is a general feature of acceptor-doped GaAs/GaAlAs heterostructures and, again, it is similar to the behavior CR in Si-MOS systems.

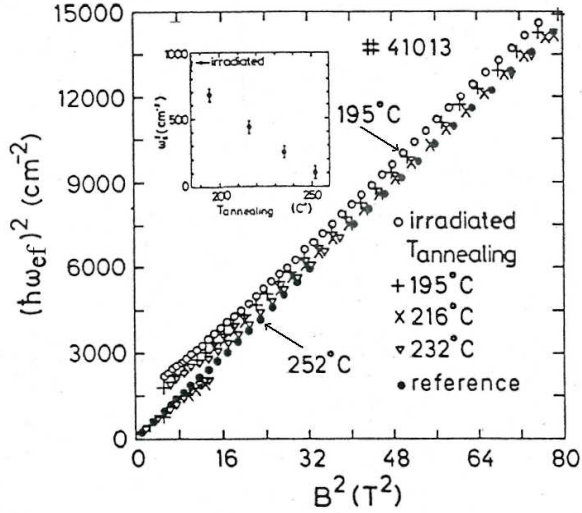

Figure 25. Effective frequency squared versus $B^2$ in a CR-type experiment showing the result of annealing process on the irradiated GaAs/GaAlAs heterostructure. In the inset, the experimental shift $\omega_s^2$ (offset squared) is plotted as a function of the annealing temperature between 195 °C and 252 °C.. After Ref. [50]

Data similar to those shown in Figs. 24 and 25 were obtained on GaAs/GaAlAs heterostructures $\delta$-doped in the well with Be acceptors. Again, two resonances were observed at lower magnetic fields: regular CR and DM. Although the introduced acceptors are not located in a very thin sheet parallel to the interface because of the atomic diffusion processes, see Ref. [53], one deals here with two different parts of the quantum well: the undoped GaAs part and the doped part. It is then understandable that the undoped part gives the normal CR excitation while the doped part gives the DM excitation. It is unclear why the irradiated samples and the $\delta$-doped samples exhibit very similar spectra. It was remarked in Ref. [50] that the shift of DM is related to some average value of the scattering strength (average localization). The following relation between the offset squared and the density of scatterers (acceptor ions) was deduced

$$\omega_s^2 = \frac{8}{m^*} V_0 \times N_{sc}, \qquad (14)$$

where $V_0$ is the depth of localizing well and $N_{sc}$ is the sheet density of scatterers (acceptor ions). This agrees with the dependence illustrated in the inset of Fig. 25. Thus the deduced phenomenological model is of an electron bound by a harmonic potential. The latter is determined by collective properties of the random scatterers (acceptor ions). The width of the confining potential is determined by the mean distance between acceptors. This model was corroborated by later theoretical considerations.

The influence of impurities on the cyclotron resonance was also investigated at the Oxford University in high mobility GaAs/GaAlAs heterostructures not intentionally doped in the well [54, 55]. In this case one sees a single resonance as a function of *B*, going continuously from the usual CR line to the shifted DM line. The reason for this behavior is that possible splittings (offsets) are smaller than the resonance half-width. Magnetic field and electron density dependences of the offsets were investigated and it was shown, at least in 10 studied samples, that the offsets decreased with increasing 2D electron density $N_s$ [55]. This is illustrated in Fig. 26. According to today's knowledge, the heterostructures described in Refs. 54 and 55 contained small quantities of residual acceptors (probably carbon atoms).

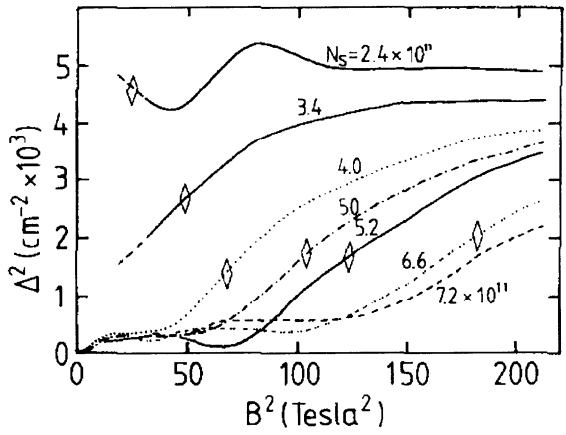

Figure 26. Offset squared versus magnetic field squared for GaAs/GaAlAs single quantum well, at a range of different electron densities $N_s$ (in cm$^{-2}$), as changed by an illumination with red light emitting diode. The diamonds mark positions of the filling factor $v = 2$. It is seen that the offsets decrease with increasing $N_s$ and, for a given $N_s$, they increase with *B*. After Ref. [55].

### 4. 2 THEORY

As to the theoretical description of the disorder modes, we will not go to its details referring the reader to original papers. The basic idea, which already emerged from the experimental work reviewed above, is that the presence of repulsive acceptor ions located near each other results in local potential minima. The latter may be approximated by parabolic wells binding the electrons, as indicated in figure 27. The main theoretical difficulty is that various wells are not identical because the acceptors are located randomly in the $\delta$-doping plane, so that one deals with a distribution of wells. However, since there exists an average distance between acceptors, one can define an average, or "most probable" well $(1/2)m^*\omega_s^2$. This average well determines, in the first approximation, an additional binding due to localization, i.e. the offset of the disorder mode. This is in fact what is theoretically predicted. Another important ingredient of the theory is the electron-electron (e-e) interaction. In the presence of the latter one can assume that the 2D electron gas, driven by the light wave, moves *rigidly* against the background impurity potential, which gives rise to an effective harmonic restoring force. This also means that internal excitations within the 2D gas are suppressed.

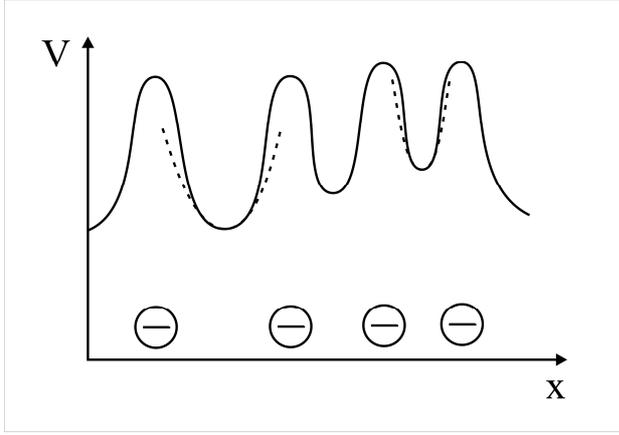

Figure 27. Disorder potential due to negatively charged acceptors in the quantum well (schematically). Potential fluctuations result in various 2D potential wells which, on average, shift upwards the electron cyclotron frequency.

To describe the effect of impurities on CR in 2D systems Zaremba [56] used general sum-rule arguments for the power absorption of incident electromagnetic radiation. The calculation gives a general expression for the DM frequency with the positive offset proportional to the number of acceptors, in agreement with Eq. (14). Under some simplifying assumptions the problem can be reduced to classical equations of motion and then also a second, low-frequency mode is obtained, as described by $\omega^2 = \omega_s^4/4\omega_c^2$. This mode has never been observed. Merkt [57] used similar assumptions to those of Ref. [56]. Separating the initial Hamiltonian into the center-of-mass part and the relative-coordinate part, it was assumed that the center-of-mass part is described by the harmonic potential with the average frequency $\tilde{\omega}_0^2 = (1/N)\sum_{i=1}^{N}\omega_i^2$, in which the sum runs over all the wells. One then obtains two optical resonances

$$\omega_\pm = \left[\tilde{\omega}_0^2 + \left(\frac{\omega_c}{2}\right)^2\right]^{1/2} \pm \frac{\omega_c}{2}, \qquad (15)$$

for the two circular light polarizations. With appropriate definitions this agrees with the previous descriptions. The harmonic frequency resulting from a potential well between two acceptors was estimated for the case in which acceptors are in the GaAlAs barrier, at a certain distance from the 2DEG. One obtains then: $\tilde{\omega}_0^2 = 4e^2/\pi\varepsilon\varepsilon_0 m^* R_{Be}^3$, when the distance $\Delta z$ between the acceptor doping sheet and the plane of the 2DEG is less than the separation $R_{Be}$ between the acceptors given by the relation $R_{Be} = N_{Be}^{-1/2}$. This relation gives a reasonable estimation of various experimental data. One important shortcoming of the theory is that Eq. (15) predicts also the lower frequency mode, never observed. In addition, the theory does not really account for the observed qualitative differences of behavior between donor-doped and acceptor-doped heterostructures. Clearly, one essential experimental difference between the two situations is that, in the standard case of 2D electrons coming from the donors in the GaAlAs barrier, the additional donors in the well are neutral while the additional acceptors are negatively charged.

## 4. 3 LATER STUDIES

Cyclotron resonance studies at the University of Hamburg were carried out on GaAs/GaAlAs heterostructures $\delta$-doped by Be acceptors. In contrast to the previous investigations, the structures were doped in the GaAlAs barrier, 2-3 nm from the interface with GaAs. Widmann et al. [58] observed regular disorder modes shifted to higher energies with respect to the CR excitations. It was found that the offset increased with increasing 2D electron density, as illustrated in Fig. 28. This behavior is opposite to that observed in Ref. [55] and shown in Fig. 26. Buth et al. [59] observed a very weak second DM and suggested a perceptible degree of the short-range order between the sites of beryllium acceptors. In Ref. [60] the authors simulated potential landscapes for electrons in random and correlated distributions of repulsive scatterers and applied it to quantum magneto-transport. It was concluded that the question of possible correlation between the sites of Be acceptors could be answered only by a direct determination of the beryllium distribution.

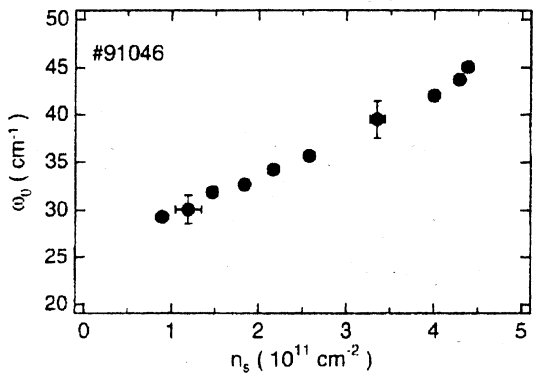

Figure 28. The offset frequency $\omega_0$ for GaAs/GaAlAs heterostructure doped in the barrier with Be acceptors ($N_{Be} = 2\times10^{10}$ cm$^{-2}$) versus 2D electron density. The observed increase is in contrast to the results shown in Fig. 26 for residual acceptors in GaAs/GaAlAs samples. After Ref. [58].

The disorder modes were also observed on GaAs/GaAlAs heterostructures doped in the well by Be acceptors, see Ref. [19]. The unique feature of these data is that they exhibit both the effects of disorder, which is by definition a multi-acceptor phenomenon, as well as the single-acceptor excitations in the form of "small resonances", see Fig. 10. In particular, it is seen from this figure that one deals with *two* disorder modes. The energy square versus $B^2$ plot for the three main peaks is shown in Fig. 29. The characteristic parallel behavior of the energies in the square-square plot is not perfect. Also, the CR and DM excitations are observed together for the whole range of accessible magnetic fields. The DM peak begins to be seen around $v = 4$, in contrast to the observations of other authors, who saw DM for much smaller filling factors (around $v \leq 2$). Two DMs were observed on three heavily doped samples.

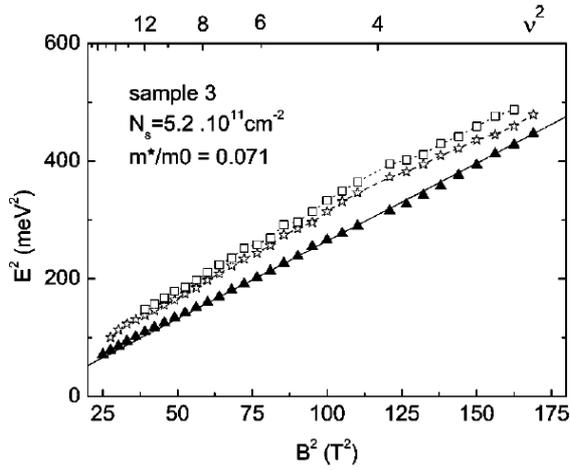

Figure 29. Cyclotron resonance and two disorder modes observed on GaAs/GaAlAs heterostructure doped in the well by Be acceptors (energy square vs. $B^2$). The straight line for CR is theoretical, the lines for DM's are drawn to guide the eye. The DMs begin to be seen at the filling factors $v \cong 4$. After Ref. [19].

The origin of multiple DM excitations is not clear. As indicated above, the standard picture of the DM frequency results from the acceptor potential fluctuations and the strong e-e interaction. There is no obvious reason why this mechanism should give more than one oscillator frequency. The two DM frequencies could result from two doping planes, one introduced by the actual $\delta$-doping in the well at 2 nm from the interface, the other resulting from the diffusion of Be atoms, probably at the GaAs/GaAlAs interface. In this case, however, the two frequencies should differ more from each other than what is observed, since the acceptor density at the interface plane should be considerably lower than that of the doping plane. We conclude that the number of DM excitations is not understood since, as remarked above, it is not even clear why in the uniformly doped samples one observes separately CR and DM modes (see Fig. 10) and why the uniformly doped and $\delta$-doped samples give very similar spectra.

The disorder modes were observed in GaAs/GaAlAs heterostructures doped in the well with *carbon* acceptors [61, 62]. The splitting between DM and CR resonances increased with the increasing number of scatterers, similarly to the results of other authors, see Fig. 25, and increased with growing 2D electron density, similarly to the results shown in Fig. 26. The authors stated that the origin of the splitting remained unclear but their results confirm in general the data taken on Be-doped samples, so the interpretation should also be similar.

It follows from the above work that C acceptors localize conduction electrons similarly to Be acceptors. This explains why samples containing probably residual C atoms (see Refs. [12, 55]) behave similarly to those containing Be atoms introduced by modulation doping.

## 4. 4 DISORDER MODES AT HIGH MAGNETIC FIELDS

One of the characteristic properties of acceptor-doped samples is that the disorder mode at high magnetic fields approaches the regular CR and appears in the form of a very narrow resonance, see Fig. 24 and Ref. [19]. The line-width of this resonance versus *B* is shown in Fig. 30. This feature is present in *all* acceptor-doped GaAs/GaAlAs heterostructures and for all doping schemes mentioned above: homogeneous, residual, doping in the well and doping in the barrier. Sigg et al. [50] observed that the very narrow resonance width at high *B* shows no significant dependence on the number of acceptors. The authors remarked: "This suggests, that at high magnetic fields the electrons are avoiding the scatterers so perfectly as if no scatterers were present." Richter et al. [51] suggested that the narrowing of the CR line arises from the fact that at high fields the electrons fall down from the magneto-acceptor levels to the free Landau levels, so that no acceptor states are occupied. This means that the acceptor states become neutral and produce no potential disorder. (We can add here that this process was proposed above under the name of "rain down", see section 2. 3. 1. However, in our opinion, after the rain down process one still deals with the negatively charged acceptor ions.)

Zaremba [56] considered a positive acceptor impurity potential and the corresponding screening electron density having a minimum at the impurity site. Using the selfconsistent Hartree calculations as a guide, see Ref. [63], he showed that for $v \leq 1$ the contact electron density can reach the zero value. When this happens, the resonance width $\Gamma$ can decrease $10^2$ times relative to the situation where the screening density is finite. That is what one observes. This result implies reduced dissipation and the resonance in this limit can be considered as a coherent cyclotron motion in the presence of a harmonic confining potential arising from the impurity background. However, the narrowing of the DM resonance in this theory is a one-acceptor effect. The situation with donor impurities is quite different since the screening electron density has a maximum at the donor site and never reaches the zero value. That is why the Si-doped systems exhibit a much broader resonance as compared to the Be-doped samples.

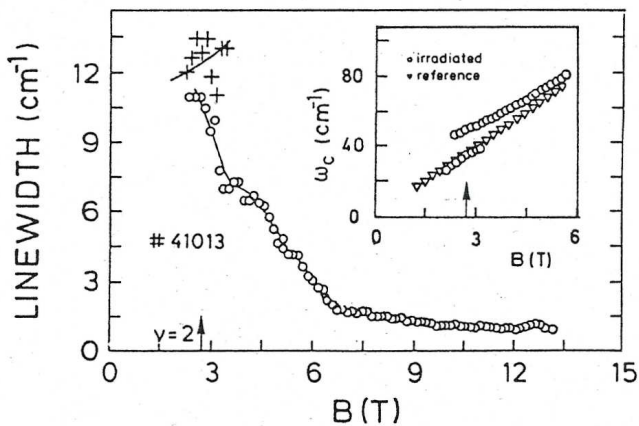

Figure 30 . Line-width of the disorder mode (empty circles) versus magnetic field for GaAs/GaAlAs heterostructure doped by electron irradiation. The inset shows resonance frequencies of the undoped reference sample (triangles), and the doped sample (circles). The filling factor $v = 2$ is indicated by the arrow. After Ref. [50].

## 5. ANTIDOTS IN A MAGNETIC FIELD

It has been remarked that antidots can influence surrounding electrons similarly to potential fluctuations and negative acceptor ions. Below we briefly trace this similarity without going into details and quoting only two typical references. Lorke et al. [64] investigated the cyclotron resonance in GaAs/GaAlAs heterostructures modulated by arrays of dots and antidots and observed a two-mode behavior similar to the one discussed above in Section 4. For the system with antidots there are two modes with the frequencies

$$\omega_{\pm} = \left[ \omega_0^2 + \left( \frac{\omega_c}{2} \right)^2 \right]^{1/2} \pm \frac{\omega_c}{2}, \qquad (16)$$

similar to the result for acceptor-doped hetrostructures, see Eq. (15). The authors interpreted this result as a manifestation of the electron localization by antidots. In a later work Suchalkin et al. [65] studied the cyclotron resonance in GaAs/GaAlAs heterostructures with randomly distributed self-organized GaInAs antidots. A strong narrowing of the CR line at increasing magnetic field was observed, as well as a switch from the cyclotron line to a higher- frequency line, again similarly to the behavior in acceptor-doped heterostructures, see Fig. 30. The higher-frequency line could be described by Eq. (16). The offset decreased with the 2D electron density $N_s$, as long as $N_s$ was inferior to the antidot density. The authors emphasized that, because of the disorder in the system, the Kohn theorem was not valid and suggested "formation of strongly correlated electron state". Thus, not only the antidot systems in a magnetic field behave similarly to the acceptor-doped ones, but also the interpretations are similar varying between the localization and correlated electron state. Still, everybody agrees that in both systems the electron-electron interaction is important.

## 6. SUMMARY AND PROSPECTS

We have reviewed properties of GaAs/GaAlAs heterostructures in which 2D electron gas in GaAs quantum wells is influenced by acceptors (mostly Be and sometimes C atoms) either residual or introduced into the vicinity of the wells or in the wells by $\delta$-doping. Magneto-optical and magneto-transport properties of such heterostructures are quite different from those doped additionally with Si donors. The review is divided into three main subjects: 1) Discrete states of conduction electrons confined to the vicinity of negatively charged acceptors, 2) Asymmetry of plateaus in the Quantum Hall Effect and the Shubnikov-de Hass effect, 3) Disorder modes in the cyclotron resonance. All three subjects comprise both theory and experiment. In the subject of discrete states we discuss the variational theory and quote experimental evidence provided by photo-magneto-luminescence, cyclotron resonance and quantum magneto-transport. Two new effects are introduced, related to the existence of repulsive electron energies above the corresponding Landau levels: the rain down effect at low electric fields and the boil off effect at high electric fields. In the subject of plateaus asymmetry it is shown that, for the majority of cases, the observations can be explained by the asymmetric density of Landau states due to the presence of acceptors. However, at small filling factors (high magnetic fields) also the rain down effect comes into play. As to the disorder modes, they originate from collective properties of charged acceptors. In some experiments both the effects of individual acceptors ("small resonances" in CR) as well as the collective acceptor effects (disorder modes) are simultaneously observed.

The weak point of the reviewed research is that it includes mostly acceptors introduced by Be atoms and investigates only GaAs/GaAlAs heterostructures. We quote some early work on Si-MOS structures, but it is doubtful that they contain acceptors; they rather contain defects or inhomogeneous surface regions acting as potential wells. Thus, an obvious extension of the existing work would be to use other acceptors (carbon atoms) and other hetrostructures to see whether the observed effects are universal or not. In the subject of discrete electron states in the vicinity of negatively charged acceptors, their existence has been convincingly demonstrated in three different experiments, but one should remark that the photo-magneto-luminescence results and the magneto-transport results were published in single papers. Clearly, an additional confirmation would be welcome. As to the disorder modes, related to collective random acceptor distributions, it is not understood why different acceptor doping regimes: - uniformly doped in the well and the barrier, doped only in the barrier, doped only in the well – result in very similar behavior. The theory of DMs is difficult and, to our knowledge, there exist very few theoretical papers treating this subject. In particular, the commonly observed dramatic narrowing of the DM resonance at high magnetic fields (low filling factors) is seriously treated in only one theoretical paper [56]. Finally, one can imagine a complete reversal of the investigated systems -- instead of studying the effect of acceptors on the 2D electron gas in n-type heterostructures one could study the effect of donors on the 2D hole gas in p-type heterostructures.

ACKNOWLEDGMENTS

We dedicate this work to the memory of Stephane Bonifacie, who greatly contributed to the subject of acceptor-doped heterostructures, both theoretically and experimentally. Stephane died suddenly of a heart attack at the age of 32. Elucidating discussions with Professor Christophe Chaubet and Dr hab. Tomasz Rusin are gratefully acknowledged.